\documentclass[aps,prb,twocolumn,floatfix]{revtex4}
\usepackage{graphicx}

\begin{document}

\title{Proteins and polymers}

\author{Jayanth R. Banavar}
\affiliation{Department of Physics, 104 Davey Lab, The
Pennsylvania State University, University Park PA 16802, USA}

\author{Trinh Xuan Hoang}
\affiliation{Institute of Physics and Electronics, Vietnamese
Academy of Science and Technology, 10 Dao Tan, Hanoi, Vietnam}

\author{Amos Maritan}
\affiliation{Dipartimento di Fisica `G. Galilei' and INFN,
Universit\`a di Padova, Via Marzolo 8, 35131 Padova, Italy}

\begin{abstract}

Proteins, chain molecules of amino acids, behave in ways which are
similar to each other yet quite distinct from standard compact
polymers. We demonstrate that the Flory theorem, derived for
polymer melts, holds for compact protein native state structures
and is not incompatible with the existence of structured building
blocks such as $\alpha$-helices and $\beta$-strands. We present a
discussion on how the notion of the thickness of a polymer chain,
besides being useful in describing a chain molecule in the
continuum limit, plays a vital role in interpolating between
conventional polymer physics and the phase of matter associated
with protein structures.

\end{abstract}


\maketitle

\newcounter{ctr}
\setcounter{ctr}{1}

Proteins are chain molecules made up of small chemical entities
called amino acids.  In spite of their small size, the diverse
physical and chemical attributes of the twenty types of naturally
occurring amino acids and the history-dependent role played by
evolution, globular proteins exhibit a range of striking common
characteristics \cite{RMP}. Traditional attempts at creating a
framework for understanding proteins using ideas from polymer
physics have been largely unsuccessful as stated by
Flory\cite{Flory}: ``Synthetic analogs of globular  proteins are
unknown. The capability  of adopting a dense  globular
configuration stabilized by self-interactions and of transforming
reversibly to the random coil are  peculiar to the chain molecules
of globular proteins alone."  The standard models of polymer
physics do not provide an explanation for why there are a
relatively small number (of order thousand) native state folds
\cite{Chothia}, why they are inevitably made up of helices and
sheets \cite{Creighton} and how these folds are adapted for
biological function especially enzymatic activity.

In this paper, we seek to bridge this apparent gap between
polymer physics and the physics of compact biomolecules.  We do
this in two complementary ways: first, we study the average
behavior of compact protein native state structures and show that,
in spite of being made up mainly of $\alpha$-helices and
$\beta$-strands, the Flory theorem derived for polymer melts
\cite{polymerbooks,Orland} holds reasonably well
for native state protein structures as well; second, we
demonstrate that the notion of an anisotropic chain of non-zero
thickness is valuable for extrapolating from conventional polymer
physics to the phase used by nature to house protein structures.

Let us begin with an analysis of protein native state structures
from the protein data bank \cite{PDB} to assess the validity of
the Flory theorem. We consider a coarse-grained description in
which each amino acid is represented by its $C^{\alpha}$ atom, the
hinges of the protein backbone. It is well known from Flory's work
in polymer physics that polymer melts or even a long compact
polymer has very interesting sub-structure
\cite{polymerbooks,Govorun,Lua}. The
basic idea is that a short labelled piece of a polymer chain from
within such a dense melt exhibits statistics (distributions and
an end-to-end distance) which are characteristic of random walk behavior.
Physically, the effective absence of any interaction
is believed to arise from the inability of the chain to
discern whether it is making contacts
with itself or with other chains.  Does the presumed
validity of the Flory theorem and the existence of Gaussian random
walk statistics for short chain segments preclude structures
built up from helices and sheets? Interestingly, it has been
suggested recently \cite{Rose} that model denatured proteins can
exhibit random coil statistics in spite of having significant
secondary structure.

Our principal results are summarized in Figures 1-5 and
demonstrate that for
compact proteins, characterized by an end-to-end distance
scaling approximately as the cube root of the protein size (see
Figure 1):

1) The Flory theorem is found to hold (Figure 2) for proteins
segments made up of more than 48 amino acids.  The existence of
secondary motifs results in an effective persistence length of this order
beyond which one obtains Gaussian statistics (Figure 3)
accompanied by random walk behavior.

2) The validity of the Flory theorem is {\em not} incompatible
with the existence of secondary motifs \cite{Lua}.

3) One can understand the crossover in Figure 2 by studying
correlation functions of the tangent and the binormal vectors
along the chain (Figures 4 and 5).

Our results vividly demonstrate that proteins exhibit properties
that are not incompatible with those of generic compact polymers.
However, as stated before, the standard models of polymer physics
do not account for the rich phase of matter associated with
protein native state structures.  In order to proceed, let us
recall that a dominant structural motif used in biomolecular
structures is the helix \cite{WC,Pauling1}. An everyday object
which, on compaction, can be coiled naturally and efficiently into
a helical shape is a garden hose or a tube \cite{MaritanNature}. A
tube can be thought of as a thick polymer, a polymer chain endowed
with a natural thickness. We will proceed to study the attributes
of a tube and its relationship with conventional descriptions of
polymers.

In the continuum, a non-zero chain thickness serves a valuable
purpose. Consider first a polymer chain of vanishing thickness in
the continuum. It is well-known \cite{polymerbooks} that the end
to end distance, $R$, of a swollen, self-avoiding chain scales
approximately as the $3/5$-th power of its length, $L$. In the
absence of any other length scale in the problem (recall that we
are dealing with a chain of zero thickness in the continuum), one
is led to a fundamental problem in simple dimensional analysis in
expressing the relationship $R \sim L^{0.6}$ -- both $R$ and $L$
have units of length and there is no other length scale in the
problem which can be used to fix the correct dimension in the
scaling relation. In order to study a chain molecule in the
continuum, the traditional approach has been to use the powerful
machinery of renormalization group theory \cite{Wilson}. A tube of
non-zero thickness circumvents this problem by providing the
required additional length scale naturally, even in the continuum.
Indeed, one may write a scaling form $R(L, b, \Delta) = L
F(L/\Delta,b/\Delta)$, where $\Delta$ is the tube thickness. The
continuum limit can be safely taken by letting $b$ go to $0$
leading to $R = L F(L/\Delta,0) \sim \Delta^{1-\nu} L^\nu$.

An interesting issue in polymer physics is the description, in the
continuum, of a closed chain with certain knot topologies. One, of
course, requires physically that the knot number be preserved in
any dynamics.  A string described by in standard continuum
approach is necessarily characterized by an infinitesimal
thickness and allows changes in the knot topology with a finite
energy cost rendering the model somewhat unphysical in this
regard. This problem is cured by the tube description. Hard
spheres have been studied for centuries and their self-avoidance
is ensured by considering all pairs of spheres and requiring that
their centers are no closer than the sphere diameter. Strikingly,
the generalization of this result to a tube entails a simple
modification of the standard pair-wise interactions \cite{BGMM}.
For each pair of points along the tube axis, one draws two circles
both passing through the two points and each one tangential to the
axis at one or the other location. One then simply requires that
none of the radii is smaller than the tube radius \cite{GM,BGMM}.
The use of many-body potentials is an essential ingredient for
describing a tube in the continuum \cite{BGMM}. The many-body
potential replaces the pairwise self-interaction potential and
ought not to be thought of as a higher order correction.

The coarse-grained flexible tube model captures two essential
ingredients of proteins  -- the space within a tube roughly allows
for the packing of the protein atoms and local steric effects are
encapsulated by constraints on the local radius of curvature; the
effects of the geometrical constraints imposed by the chemistry of
backbone hydrogen bonds are represented by the inherent anisotropy
of a tube (a  tube, when discretized, may be imagined to be a chain of
discs).  The generic compact polymer phase arises for long tubes with a
thickness much smaller than the range of attractive interactions
promoting compaction.

Recent work \cite{HoangPNAS} has shown that the low energy conformations
adopted by tube-like polymers with certain constraints on symmetry and
geometry are made up of helices and sheets akin to marginally compact
protein secondary structures.
For classes of short homopolymers characterized by generic
geometrical constraints arising from backbone hydrogen bonds and
sterics and with mild variations in their overall hydrophobicity
and local curvature energy penalty parameters, one obtains a free
energy landscape\cite{HoangPNAS}, determined by geometry and
symmetry, with multiple minima corresponding to the menu of folds.
We have generated a thousand structures with low energies of a
homopolymer of length $N=48$. The structures are local energy
minima in simulated annealing simulations. A refined set of about
320 protein-like structures is obtained by choosing only those
that are marginally compact ($7.6\AA < R_g < 12\AA$) and have a
sufficient amount of secondary structure content (the fraction of
residues participating in either a helix or a sheet is larger than
60\% of the total number of residues). 
Strikingly, Figure 6a shows that the behavior of short segments of real
proteins and the model structures are qualitatively similar to each other.
The deviation from Gaussian behavior in both cases is due to the presence
of secondary structures, whose characteristic length scale is smaller for
the model structures than for real proteins. Interestingly, even for
relatively short segment lengths (l = 8, 12) in the model structures, one
observes statistical behavior somewhat similar to that of Gaussian chains
(Figure 6b) along with significant deviations, most notably a peak due to
the presence of the secondary structures.  Due to the limited chain length
that one can reliably study in the model we are not able to observe the
crossover to the regime predicted by Flory.

In summary, we have shown that there is a natural bridge, provided
by the chain thickness, between polymer physics and the physics of
biomolecular structures.  The thickness provides a physically
motivated cut-off length scale which allows for a well-defined
continuum limit. The Flory theorem is found to hold for proteins
in spite of the structured building blocks of protein native state
structures. Our results suggest that the powerful arsenal of
techniques of polymer physics can be brought to bear on the
protein problem and conversely, the notion that chain molecules
are inherently anisotropic and have a non-zero thickness provide a
new perspective in the field of polymer physics.

This work was supported by PRIN 2003, INFN, NASA, NSF IGERT grant
DGE-9987589, NSF MRSEC at Penn State, and the NSC of Vietnam (grant 
No. 410704).

\newpage

\section*{Figure Captions}

\begin{description}
\item[Fig. 1.]Log-log
plot of the radius of gyration $R_g$ of a set of 700 proteins versus their
length $L$ or the number of constituent amino acids. The proteins
used in our study were selected based on several criteria: the
sequences chosen have less than 50\% overlap with each other, the
structures have been obtained with high resolution X-ray
diffraction and the proteins are substantially compact without
dangling ends so that $R_g/L^{1/3} \leq 3.02\AA$. The straight line
has a slope of $1/3$ as a guide to the eye.
\item[Fig. 2.]Log-log
plot of the end to end distance $R$ versus $l$ for protein
segments. The plot was obtained by averaging over all segments of
length $l$ selected from the data set depicted in Figure 1. For a
given $l$, $R$ was determined as an average over all segments of
that length in proteins whose lengths are greater than $l^{3/2}$,
in order to avoid finite size effects \cite{Lua}. The error bars
are of the order of the size of the symbols. Note the plateau
which indicates that $R$ is only slowly increasing with $l$ around
24. For values of $l$ larger than 48, we find that $R \sim
l^{1/2}$.
\item[Fig. 3.]Statistics of the end-to-end distance of segments of
proteins of length $l$. For $l=48$, 64 and 80, the distributions
show a nice collapse to the form expected for Gaussian statistics:
the solid line denotes the function
$P(x)=\frac{1}{\sigma^3}\sqrt{\frac{2}{\pi
l}}\,x^2\exp(-\frac{x^2}{2\sigma^2})$, where $\sigma=2.164 \AA $ .
For $l=16$, where the presence of secondary motifs play a major
role, the distribution is qualitatively different from the other
sizes and exhibits a peak arising from the presence of
$\alpha$-helices.
\item[Fig. 4.]Plot of
the tangent-tangent and binormal-binormal correlation functions
along the protein sequence derived from our data set. The tangent
vector at location $i$ is defined as an unit vector pointing along
the line joining the positions of the $i-1$-th and the $i+1$-th
amino acids.
The normal vector is defined by joining the $i$-th location to the
center of the circle drawn through three amino acid
($i-1$,$i$,$i+1$) locations. The binormal is perpendicular to the
plane defined by the tangent and the normal. Note that: a) the
negative tangent-tangent correlation at sequence separation $k$
around 13 corresponds to a turning back,  on average, of the chain
direction and is related to the cross-over shown in Figure 2; b)
the binormal-binormal correlation remains non-zero for large
separations.
\item[Fig. 5.]Histogram of the magnitudes of the average tangent and
binormal vectors for each protein in our data set. For each
protein, we measured the magnitude as $\frac {1} {N}
\mid\Sigma_{i=1}^N \vec{v}_i\mid$, where $\vec{v}_i$ is either the
unit tangent or the unit binormal vector at location $i$ and $N$
is the number of such vectors for a given protein. For comparison,
a histogram of the magnitudes of the average of randomly oriented
vectors is shown as the shaded histogram. (Here $\vec{v}_i$ was
selected to be a randomly oriented unit vector.) Note that several
proteins have a significant non-zero mean binormal vector due to
the presence of $\alpha$-helices.
\item[Fig. 6.](a) Statistics of the end-to-end distance of
segments of length $l=6$ taken from model protein structures 
\cite{HoangPNAS} and from PDB structures. The peak in the
distributions arises from the presence of $\alpha$-helices.
(b) Same as Figure 3 but for segments of the model structures
of lengths $l=8$ and $l=12$. The fits to the
Gaussian form given in the caption of Figure 3 yield
$\sigma=2.61\AA$ for $l=8$ and $\sigma=2.08\AA$ for $l=12$.
\end{description}

\clearpage

\begin{figure}
\centerline{\includegraphics[width=3.2in]{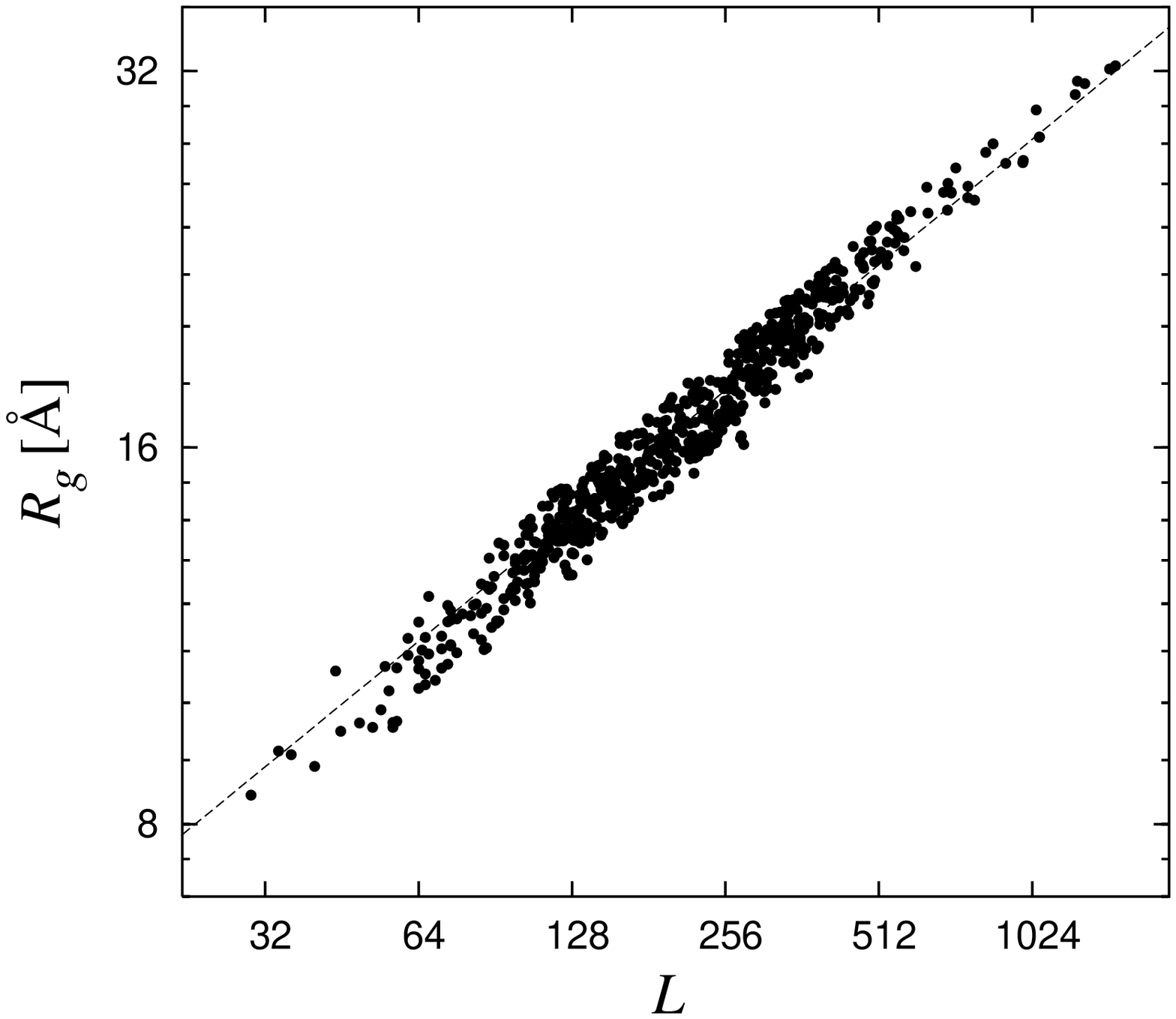}}
\caption{}
\end{figure}

\begin{figure}
\centerline{\includegraphics[width=3.2in]{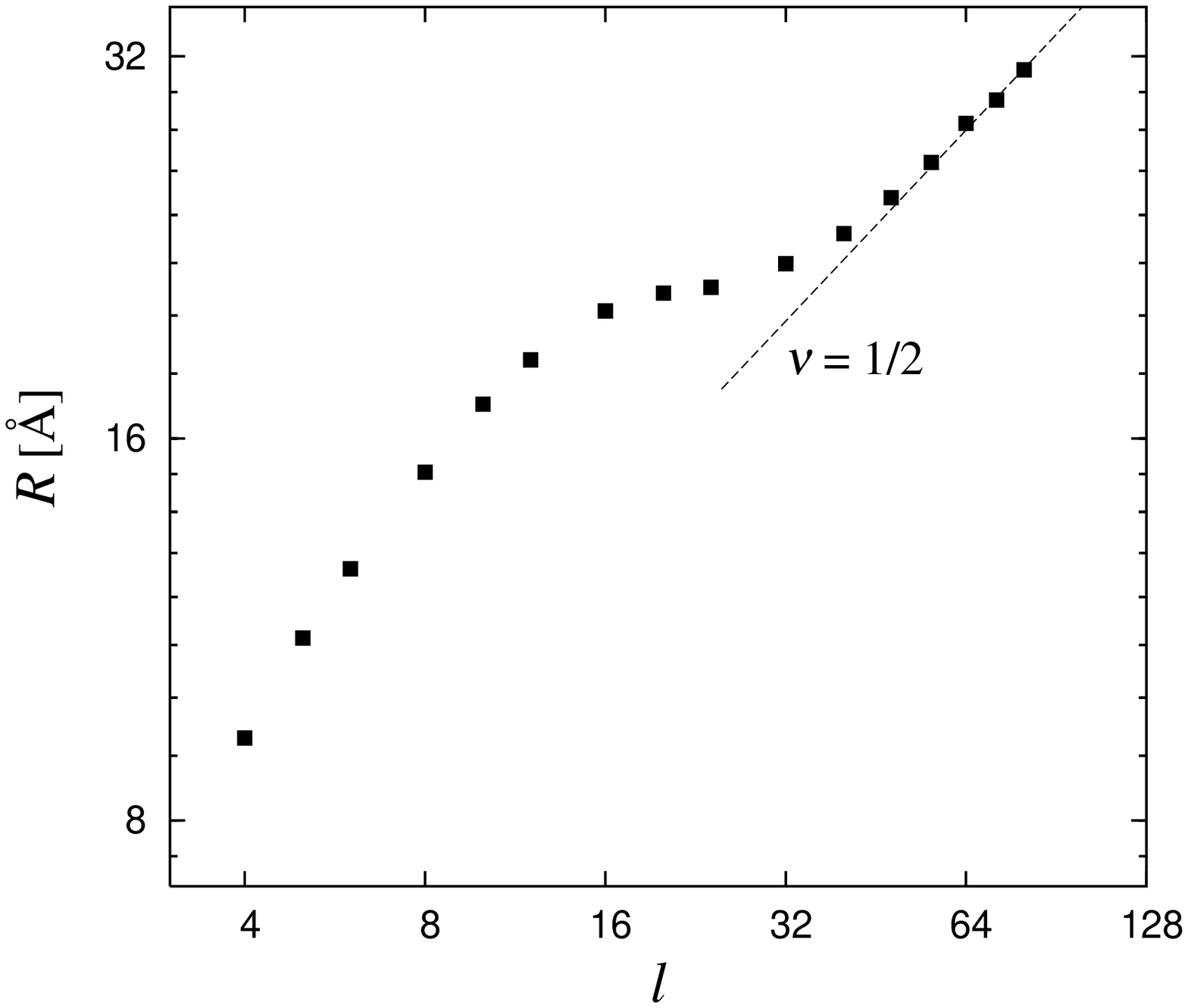}}
\caption{}
\end{figure}

\begin{figure}
\centerline{\includegraphics[width=3.2in]{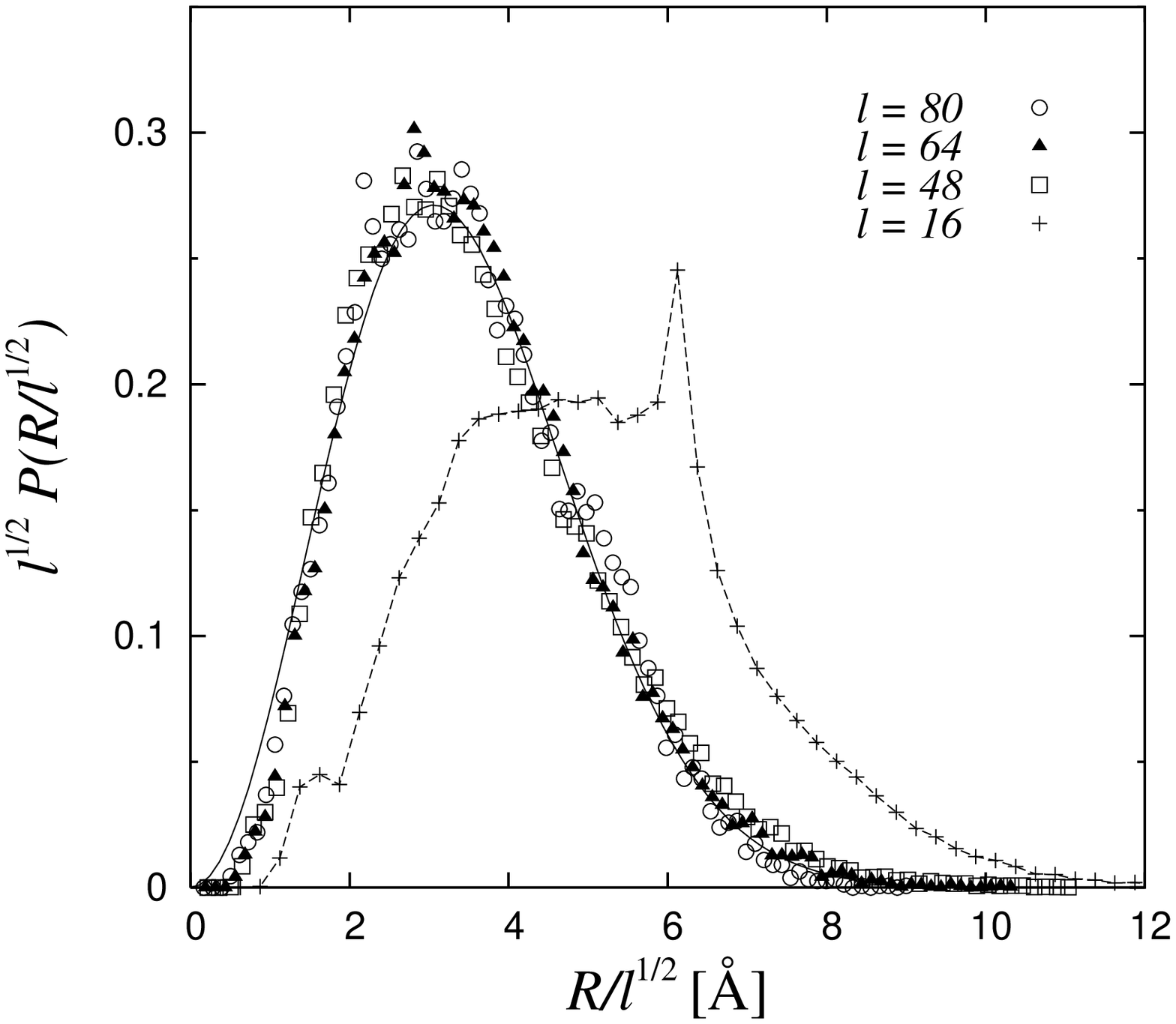}}
\caption{}
\end{figure}

\begin{figure}
\centerline{\includegraphics[width=3.2in]{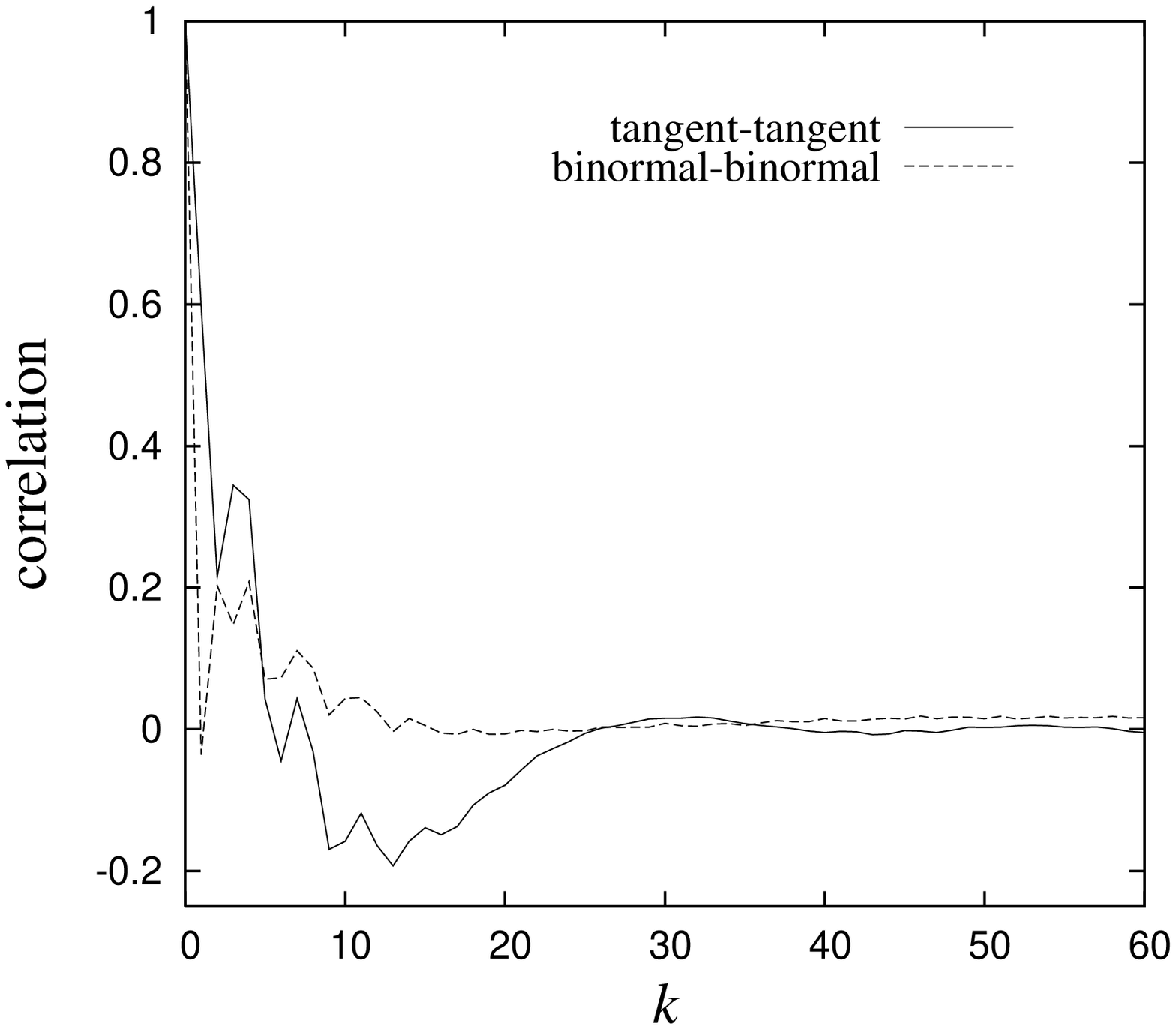}}
\caption{}
\end{figure}

\begin{figure}
\centerline{\includegraphics[width=3.2in]{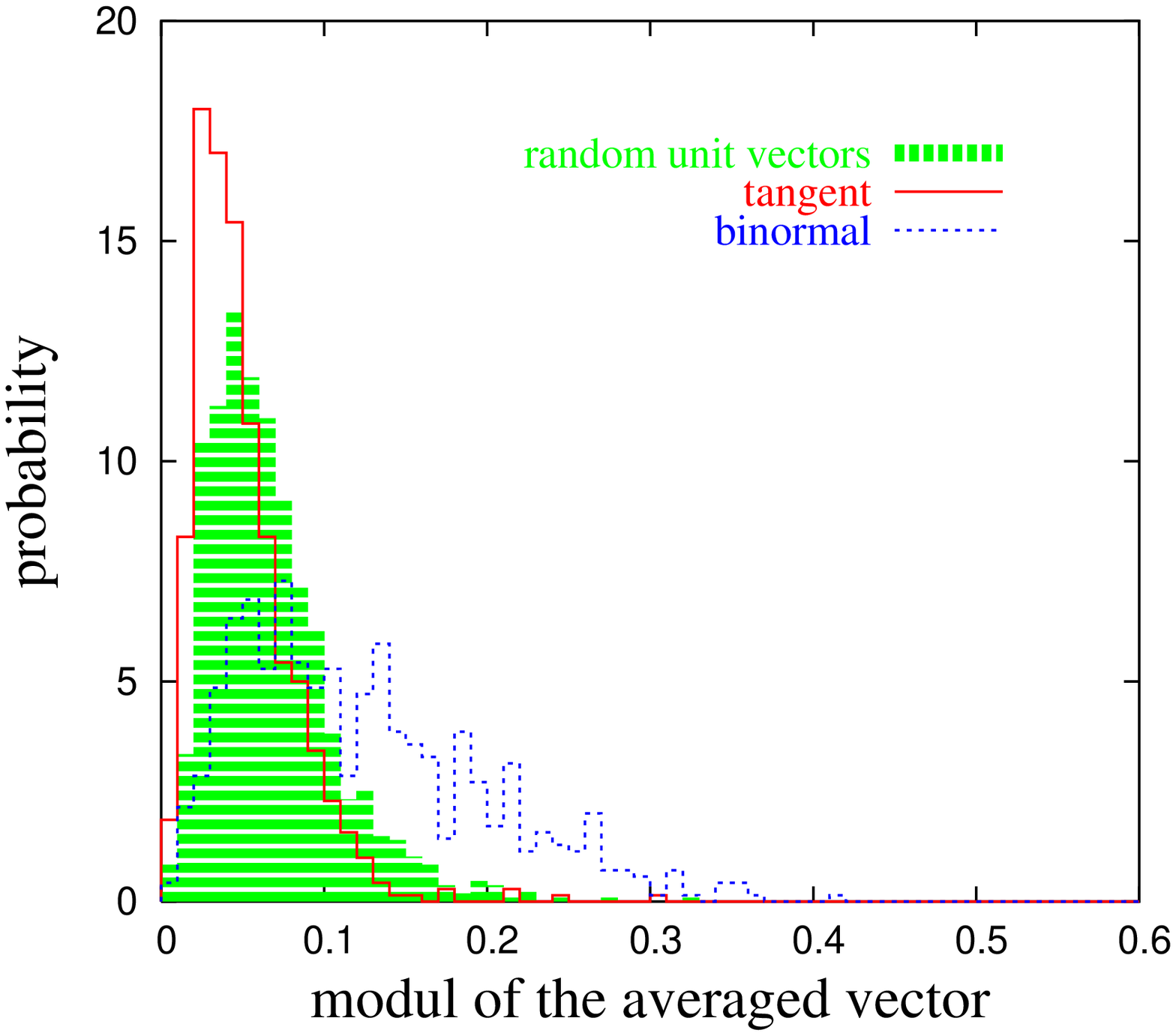}}
\caption{}
\end{figure}

\begin{figure}
\centerline{\includegraphics[width=3.2in]{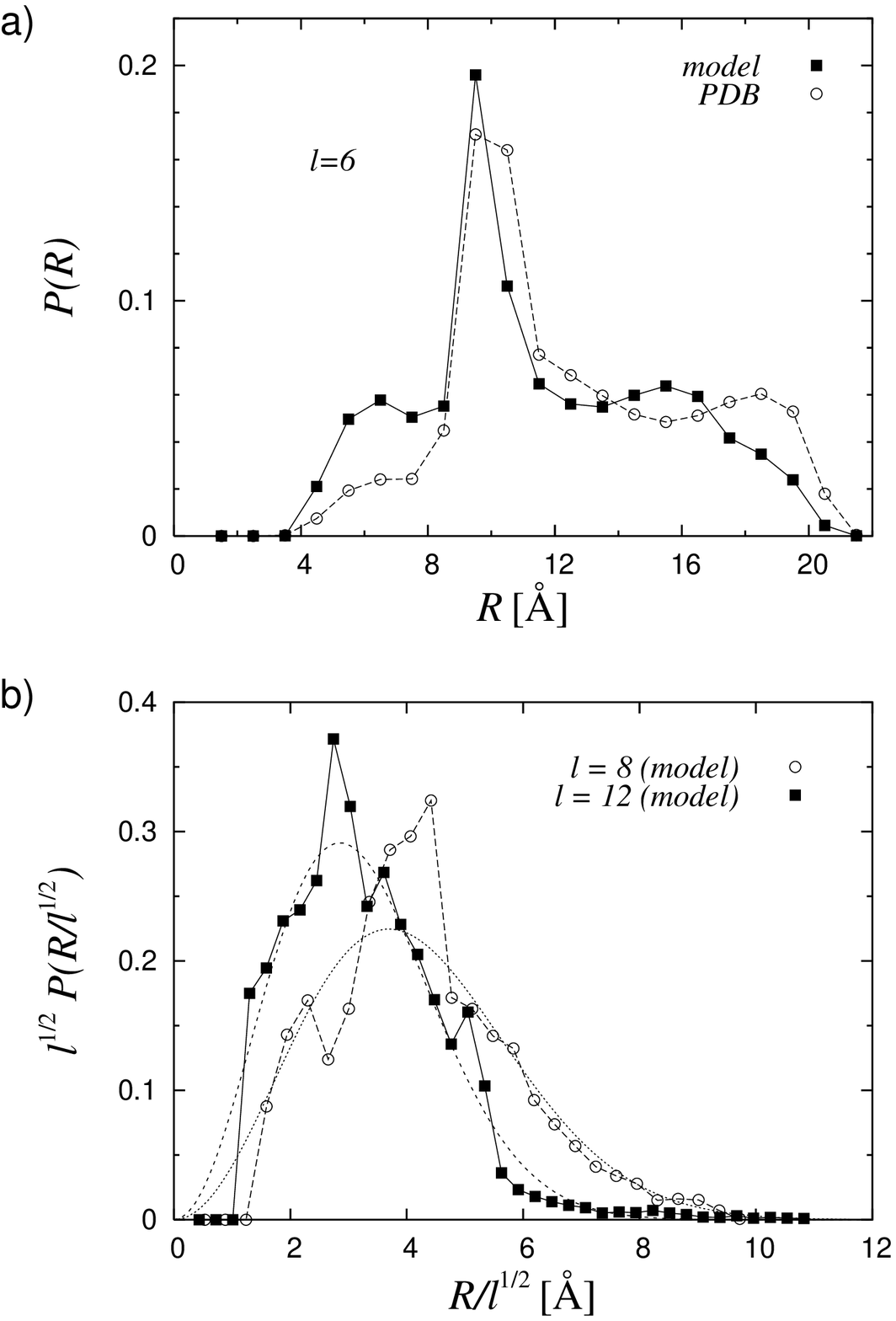}}
\caption{ }
\end{figure}

\end{document}